\def\systema{SrMnO$_3$\/ }
\def\systemb{PrMnO$_3$\/ }
\def\systemc{Pr$_{0.75}$Sr$_{0.25}$MnO$_3$\/ }
\def\systemd{Pr$_{1-x}$Sr$_{x}$MnO$_3$\/ }

\documentclass[singlespacing]{elsart}
\usepackage{graphics}
\usepackage{amssymb}
\begin{document}
\begin{frontmatter}
\title{Half Metallicity in \systemc: A first Principle study.}
\author[]{Monodeep Chakraborty}, 
\ead{monodeep@iopb.res.in; Tel:+91-674-2301058; Fax:+91-674-2300142}
\author[]{Prabir Pal,}
\author[]{Biju Raja Sekhar}
\address[]{Institute of Physics, Sachivalaya Marg, Bhubaneswar 751 005,
India.}

\begin{abstract}{In this communication we present a first principle study
of \systemd with $x = 0.25$. While the parent compounds of this system are 
antiferromagnetic insulators with different structural and magnetic ground
states, the $x = 0.25$ is in the colossal magnetoresistance regime of the
\systemd phase diagram\cite{martin1}. Our band structure calculations for
the end-point compounds matches well with the existing theoretical and
experimental results\cite{martin1,ravi}. Interestingly, our calculations
show that the \systemc has a half-metallic character with a huge band gap
of $2.8$ eV in the minority band. We believe this result would fuel
further interest in some of these special compositions of colossal
magnetoresistive manganites as they could be potential candidates for
spintronic devices. We discuss the half-metallicity of the \systemc in the
light of changes in the orbital hybridization as a result of Sr doping in
\systemb. Further, we highlight the importance of half-metallicity for a
consolidated understanding of colossal magnetoresistance effect.}
\end{abstract}    
 
\end{frontmatter}

\section{Introduction}  

The colossal magnetoresistance (CMR) materials have attracted a lot of
attention of the condensed matter community owing to their spectacular
insulator-metal transition with magnetic field. Ferromagnetic
A$_{1-x}$B$_{x}$MnO$_{3}$ (A = rare earth, B = alkaline earth) exhibits
CMR properties at particular concentrations of $x$ in their respective
phase diagrams. Half-Metallicity (HM) has been observed in a few of these
compounds both theoretically and experimentally\cite{pickett,park}. In
case of half-metals one of the spin bands (generally the majority band) is
conducting whereas the other band (generally the minority band) is
insulating at the Fermi level ($E_F$). This facilitates 100\% spin
polarization. This property of the half-metals make them potential
candidates for application in spintronic devises and magnetic sensors. The
CMR effect along with high spin polarization add to the technological
importance of the CMR manganites. Apart from their great potential in
technology, the strong interplay of the spin, orbital and charge degrees
of freedom of the charge carriers involved in this insulator-metal
transition, holds out a promise for rich physics.

In this paper we have done a first principle Tight Binding- Linearized
Muffin Tin Orbital (TB-LMTO)\cite{ander1,ander2} calculation of the
end-point compositions of \systemd and with $x = 0.25$ doping. For
\systema we have done the calculation with local spin density
approximation (LSDA). For \systemb and \systemc we had to incorporate the
electron-electron correlation (LSDA+U) to account for the band gap 
in \systemb and to match our results with the available spectroscopic
data. Moreover, the charge and orbital order observed in doped manganites
also merits a LSDA+U treatment in order to account for the intra-shell (d
and f) Coulomb interaction\cite{Anisimov}. All the three calculations have
been done with Vosko-Ceperley-Alder parametrization for the exchange
correlation energy and potential. We have included Langreth-Mehl-Hu
gradient corrections to the exchange correlation. The k-mesh used for all
these self-consistent calculations was 10$\times$10$\times$10. Although,
\systema can take both cubic as well as hexagonal structures\cite{ravi},
here we have considered only the cubic (distorted) polymorph of this
perovskite since our main motivation is to study the HM in \systemc. We
have also compared the band structure results of \systemc with 
the end-point compounds. Crystal structure of the \systemb system is taken
from a published neutron diffraction data. For \systemc we have taken the
same structure as \systemb with one Pr atom replaced by a di-valent Sr. 

Electron-lattice coupling (ELC) has a very important role in the physics
of manganites. They show up in two ways. First is the so called "tolerance
factor"\cite{millis} involving the static effect of crystal structure on
electron hopping, which has a direct effect on conductivity. The atomic
size difference between the rare-earth atoms and the divalent dopants
results in an internal stress which effects the Mn-O-Mn bonds. The
electron hopping between the Mn sites is inversely proportional to the
compression of the Mn-O-Mn bonds. This type of ELC of \systemb has been
taken in to account in our calculations. The importance of different
Jahn-Teller modes and polarons in accounting the proper insulating A-type
antiferromagnetic (AFM) ground state of undoped LaMnO$_3$ has been dealt
by several groups\cite{pickett,satpathy,hamada}. The second type of ELC is
the dynamic ELC which couples the lattice vibration (phonons) with
electronic degrees of freedom. Since TB-LMTO calculations are based on
adiabatic approximation which decouples the electronic and the lattice
degrees of freedom\cite{millis}, accounting for the dynamic ELC is beyond
the scope of this work. 

Band structure calculations can provide only a qualitative description of
any system as the structural complications that exist in real systems are
difficult to accommodate in a calculation. Although, we assume proper
stoichiometry, cation vacancy and oxygen-non stoichiometry are a common in
real systems. Again in \systemc we have assumed the crystal structure of
\systemb. The effect of Pr/Sr disorder and local strains and relaxations
have not been taken into account. But still our prediction of HM in
\systemc is robust enough to merit attention.

\section{\systema}
 
\begin{figure}
\centering \resizebox{12cm}{12cm}{\includegraphics{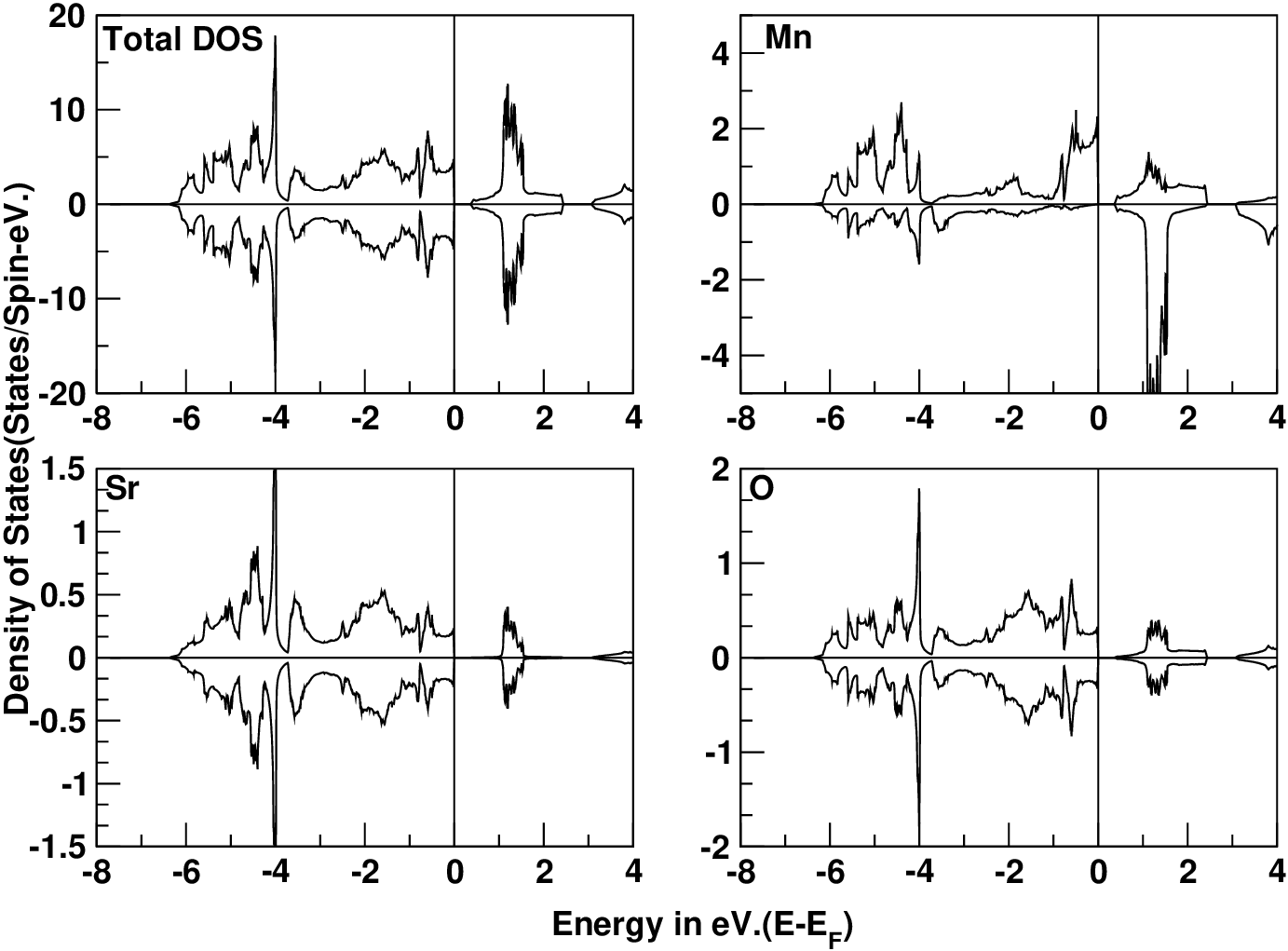}}
\caption{The spin resolved total and site projected density of states of
\systema. The top left panel depicts the total density of states for the
majority and minority band. The band gap is $0.34$ eV. The top right panel
shows the spin resolved Mn PDOS. The bottom left panel shows the Sr PDOS
and the bottom right shows the O PDOS.}  
\label{fig1}
\end{figure}

Cubic \systema has a G-type AFM ground state with a lattice
parameter\cite{ravi} of $3.824$ $A$. Our calculations with different
magnetic configurations have shown that this G-type AFM state turns out be
the most stable, in agreement with earlier reports\cite{ravi}. In this
work we have used only the G-type AFM structure. Results of our TB-LMTO 
(LSDA) calculations are shown in fig.1. The total and site projected
density of states (PDOS) clearly reveals a band gap of $0.34$ eV
which is similar to that obtained by other band structure
calculations\cite{ravi} on this system. The near $E_F$ states are
dominated by Mn $3d$ and O $2p$. The near $E_F$ valance band shows
a strong Mn $3d$ - O $2p$ hybridization. Although, Sr states have a very
little presence near $E_F$, they appear to be hybridized with the
significant O states throughout the valence band. It is clear from the
figure that the conduction states are dominated mainly by Mn. We have
estimated the magnetic moment of individual Mn atom to be $2.48$
$\mu_{\beta}$ which is again similar to $2.47$ $\mu_{\beta}$ obtained by
calculations using the Vienna Ab-Initio Simulation Package 
(VASP)\cite{ravi}. This value of the magnetic moment, which is lower
compared to that of free Mn$^{4+}$ ion (3 $\mu_{\beta}$, neglecting 
the orbital contribution), indicate the strong hybridization of the Mn and
O states in cubic \systema. It should be noted that the shape of the O
PDOS and the Mn PDOS near E$_F$ have striking similarities, which hints to
a strong covalent bonding. This inference is supported by the Crystal
orbital Hamiltonian population (COPH) analysis done by Rune Sondena et
al. \cite{ravi}.

\section{\systemb}

\begin{figure}
\centering \resizebox{12cm}{12cm}{\includegraphics{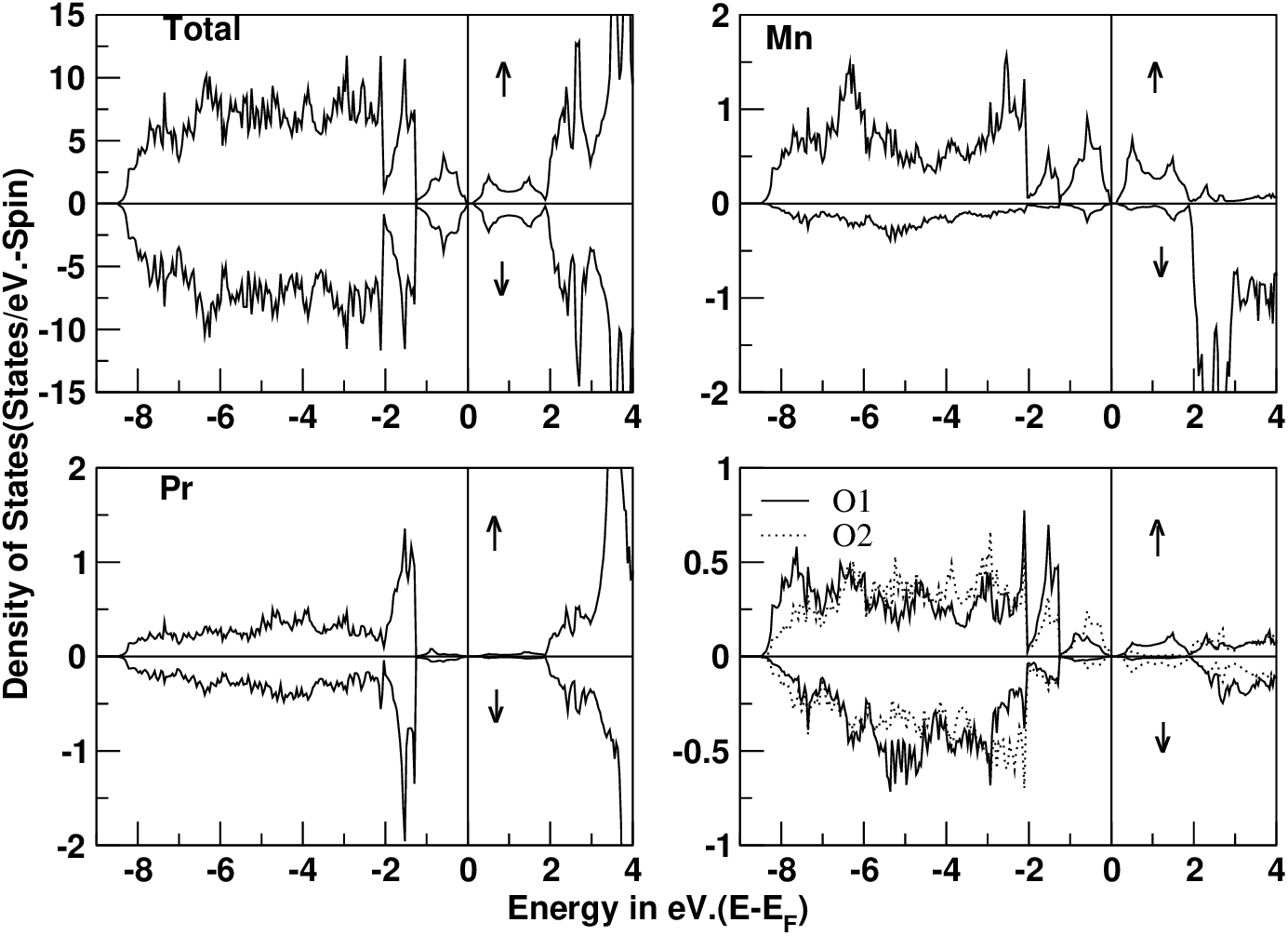}}
\caption{The spin resolved total and site projected density of states of
\systemb. The top left panel depicts the total density of states for the
majority and minority band. The band gap is $0.11$ eV. The top right panel
shows the spin resolved Mn PDOS. The bottom left panel shows the Pr PDOS
and the bottom right shows the O1/O2 PDOS.}
\label{fig2}
\end{figure}

LSDA$+$U band-structure calculations for \systemb were done for the A-type
AFM phase using the crystal structure (space group: Pbnm(62)) taken from a
neutron diffraction result\cite{zirak} published earlier. The LSDA$+$U
method was employed mainly to account for the strong electron correlation
which is behind the insulating nature of this 
material\cite{martin1}. Unlike in LaMnO$_{3}$ where simple LSDA can
reproduce the band gap, in \systemb the LSDA$+$U treatment is 
essential to derive a realistic value of the gap. In our calculation, we
have taken the exchange term $J$ and the correlation term $U$ for Pr $4f$
state to be $0.95$ eV and $7$ eV respectively and those for Mn $3d$ to be
$0.87$ eV and $4$ eV respectively. 

Fig. 2 shows the spin resolved total and site projected density of states
of \systemb. Our LSDA$+$U calculation shows an insulating A-type AFM
ground state for \systemb. We have estimated the band gap to be $0.11$
eV. Here, the electron-electron correlation that has been incorporated,
was found to be crucially important for obtaining the band gap. A simple
LSDA calculation of this system does not give an insulating ground
state. Further, the LSDA calculation results in the Pr $4f$ states
appearing very close to the $E_F$, contrary to the photoemission
spectroscopic results showing these states to be $2$ eV below the 
E$_F$. LSDA+$U$ calculation enables us to fix this problem to certain
extent. The Pr states have hardly any presence near E$_F$. The near E$_F$
PDOS of Mn and O1/O2 makes an interesting study. Though, the Mn and O1/O2
states are hybridized near E$_F$, here this hybridization is not as strong
as in the cubic SrMnO$_3$. Moreover, in this system, the hybridization
between Mn and O2 is certainly stronger than that between Mn and O1, owing
to greater physical proximity of Mn with O2 than O1. Here, the degree of
covalency of Mn-O2 bond is certainly less than that we saw in cubic
SrMnO$_3$. 

Another interesting observation is that, the hybridization of Mn 3d and
the O1/O2 (2p) states are clearly spin dependent. While there is a
considerable hybridization in the majority band, both types of oxygen
atoms hardly have any significant weight in the minority band indicting a
strong spin dependence to the Mn-O1/O2 hybridization. Here the magnetic
moment of individual Mn atom in \systemb is $3.94$ $\mu_{\beta}$ which is
comparable to the magnetic moment of free Mn$^{3+}$, neglecting the
orbital contribution. This suggests a strong atomic like moment of Mn in
\systemb and a relatively weak hybridization compared to that in 
\systema. The magnetic moment of O1 atom is $\approx 0.035$ $\mu_{\beta}$
and that of O2 atom is $\approx 0.05$  $\mu_{\beta}$. The relatively
higher magnetic moment of O2 atom could be due to its proximity with the
Mn atom.

\section{\systemc and Half-Metallicity}

\begin{figure}
\centering \resizebox{12cm}{12cm}{\includegraphics{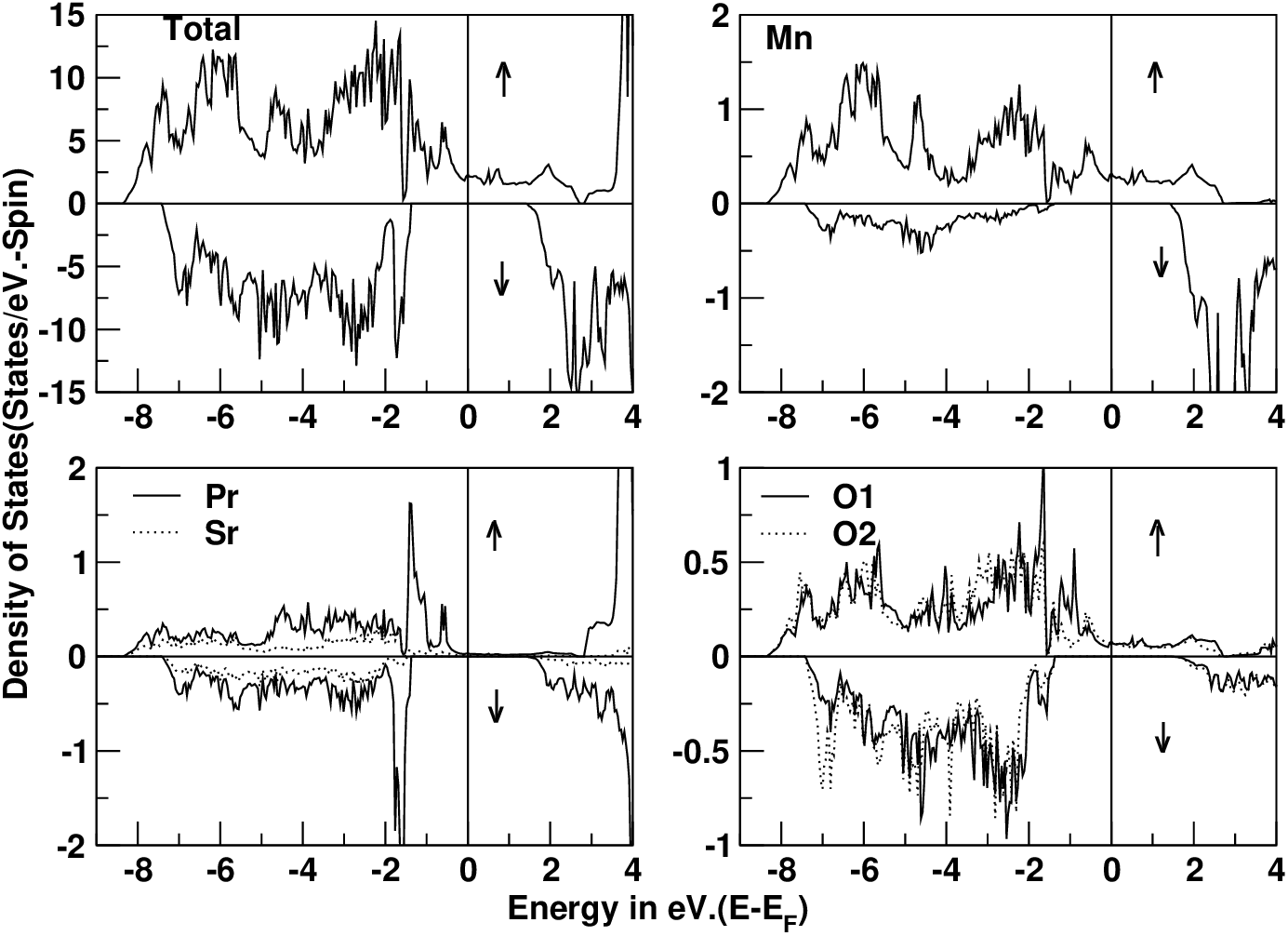}}
\caption{The spin resolved total and site projected density of states of
\systemc. The top left panel depicts the total density of states for the
majority and minority band. The minority band gap is $2.8$ eV. The top 
right panel shows the spin resolved Mn PDOS. The bottom left panel shows 
the Pr/Sr PDOS and the bottom right shows the O1/O2 PDOS.} 
\label{fig3}
\end{figure}

The band-structure calculation (LSDA$+$U) on \systemc were done for the 
ferromagnetic phase. The crystal structure that we have used for \systemc 
super-cell calculation is that of the parent compound PrMnO$_{3}$. We have 
substituted one of the Pr atoms with a Sr atom in the four formula unit
cell. The exchange term $J$ and the correlation term $U$ for Pr $4f$ and
Mn $3d$ are the same as those used for \systemb. In fig. \ref{fig3} we
present the calculated total and site projected density of states of
Pr$_{0.75}$Sr$_{0.25}$MnO$_3$. We have chosen this composition in order to
study the ferromagnetic metallic ground state of \systemd at this
doping. Also, this particular composition is easy to handle in the
LSDA$+$U. A comparison of our results with the results from 
photoemission experiments on a close composition in ferromagnetic metallic 
phase can be found elsewhere\cite{pal}. There is a qualitative matching
between our band structure results and the spectroscopic data. The most
prominent effect of substituting one of the Pr with Sr is the appearance
of a finite DOS at $E_F$ in the majority band of the total spin polarized
DOS, while the minority band display a wide gap insulating behavior. The
band gap in the minority band is 2.8 eV. Appearance of this metallicity
upon 25\% Sr doping is consistent with the phase diagram reported
earlier\cite{martin1}. We will discuss the half-metallicity upon Sr doping
by comparing these results with those from the end-point compounds, 
especially with \systemb. 

Since our focus is on HM, we will concentrate on the states near
E$_F$. The Pr/Sr states have hardly any role near E$_F$. The Sr PDOS in
\systema, though not significant, has a larger band width in the valance
band compared to that found in \systemc. This could be because of the Sr
acting more like an impurity atom. Pr PDOS in the majority band shows a
certain degree of hybridization with the O1 atom near $1$ eV below the
$E_F$. The near E$_F$ states here also are dominated by the Mn and O. From
the Mn and O PDOS, it is quite clear that there is a strong hybridization 
between the Mn $3d$ and the O $2p$ states which is spin dependent. Here
both the O1 and O2 hybridization with the Mn are stronger than the undoped
\systemb in the majority band. The O1 $2p$ shows the same degree of
hybridization at E$_F$ as that of O2 $2p$ states, which was not the case
in \systemb. We attribute this significant change to Sr doping. Comparing
the shape of O1 and O2 with those of the Mn sates in the majority band, we
conjecture that Sr doping in \systemb increases the degree of covalency
between the Mn $3d$ and O1/O2 $2p$ states. Consequently, the gap in the
majority band is filled up and gives rise to metallicity. Whereas, in the
minority band both the Mn and O1/O2 states are pushed further apart from
the E$_F$ giving rise to a band gap of $2.8$ eV. The calculated 
magnetic moment of individual Mn atom in \systemc is $3.88$ $\mu_{\beta}$,
which is less than that of Mn atom in \systemb, again suggesting a
increased hybridization activity. Simultaneously, the magnetic moment of
O1 and O2 in \systemc also increases slightly (by $\approx 0.01$
$\mu_{\beta}$) validating our conjecture of increased hybridization
activity upon Sr doping. The individual Mn atoms of \systemc loose
approximately 0.05e charge compared to the Mn atoms of \systemb. This is
due to the majority e$_g$ electrons which fill up the band gap in the
majority band as a result of the hybridization of Mn with O states. The
total magnetic moment of the Super Cell is 15 $\mu_{\beta}$. The integral
magnetic moment is the signature of the resulting HM. The magnetic moment
of Mn$_{Pr-Pr}$ atom is 3.886 $\mu_{\beta}$ and that of Mn$_{Pr-Sr}$ is
3.878 $\mu_{\beta}$. The difference in charge between the two types of Mn
are 0.002e. This shows that the difference in charge and magnetic moment
between the two inequivalent Mn atoms are insignificant. There is hardly
any difference between the two types of Mn sites in their spin projected
PDOS, with both types of Mn atoms contributing to the near E$_F$ minority
band. Pickett et. al. \cite{pickett} have tried to explain the HM 
as an effect of the A/B local environment disorder, creating a variation
in the Mn $d$ site energy which in turn induces localization effects in
the near E$_F$ minority band making it non-conducting. This does not seem
to be the case from our study. We think the spin-dependent hybridization
of Mn $3d$ and O $2p$ which was present in the \systemb is further
strengthened upon hole (Sr) doping and this is responsible for HM in
\systemc.

The HM character also can account for the high resistivity at zero
field. Since only single spin band can participate in the conductivity
process, electron hopping between ferromagnetic regions with opposite
directions of magnetization become negligible leading to high
resistivity. Here, the scattering process will not randomize the direction
of propagation of the electron as would have been the case for random
potential (spin) arrangement. Here the electron would suffer a barrier
reflection due to the ferromagnetic regions of opposite
magnetization. This effect is further accentuated in case of HM (compared
to other systems like magnetic multilayers), as there is no minority
conduction. When a magnetic field is applied forcing the different
ferromagnetic regions to align along the magnetic field, there
appears a sharp drop in resistivity. This strong insulating behaviour at
zero field and subsequent melting of different ferromagnetic regions on
application of magnetic field could contribute substantially to the large
negative magnetoresistance.

\section{Conclusions}
We have studied the \systemd with $x = 0.25$ using a first principle band 
structure calculation method of LSDA$+$U. Our calculations show that the 
CMR system \systemc has a half-metallic character with a band gap of $2.8$ 
eV in the minority band. We have compared the band structure of this 
compound with that of its parent compositions. Also, our results for the 
parent compounds match well with the existing theoretical and experimental 
results. We have discussed the half-metallicity of \systemc in the light 
of changes in spin-dependent hybridization of Mn $3d$ and O $2p$ upon hole 
doping. We have also highlighted the importance of half-metallicity for a 
consolidated understanding of CMR effect.

One of the authors (M.C) would like to acknowledge the helpful discussions
with Eva Pavarini (Forschungszentrum Juelich) and Manuel Richter (IFW
Dresden).

\end{document}